\documentclass[sigconf]{acmart}
\usepackage{array}     
\usepackage{ragged2e}  
\newcolumntype{L}[1]{>{\RaggedRight\arraybackslash}p{#1}}

\AtBeginDocument{%
  }

\setcopyright{acmlicensed}
\copyrightyear{2018}
\acmYear{2018}
\acmDOI{XXXXXXX.XXXXXXX}
\acmConference[Conference acronym 'XX]{Make sure to enter the correct
  conference title from your rights confirmation email}{June 03--05,
  2018}{Woodstock, NY}
\acmISBN{978-1-4503-XXXX-X/2018/06}

\copyrightyear{2025}
\acmYear{2025}
\setcopyright{rightsretained}
\acmConference[CSCW Companion '25]{Companion of the Computer-Supported Cooperative Work and Social Computing}{October 18--22, 2025}{Bergen, Norway}
\acmBooktitle{Companion of the Computer-Supported Cooperative Work and Social Computing (CSCW Companion '25), October 18--22, 2025, Bergen, Norway}\acmDOI{10.1145/3715070.3749235}
\acmISBN{979-8-4007-1480-1/2025/10}

\begin{document}

\title{Octo's Heartland: Supporting Children with Congenital Heart Disease through Digital Health Education}

\author{Irene Zeng}
\affiliation{%
  \institution{University of Minnesota, Twin Cities
  Department of Design Innovation}
  \city{Minneapolis}
  \state{Minnesota}
  \country{USA}
}
\email{zeng0202@umn.edu}

\author{Neda Barbazi}
\affiliation{%
  \institution{University of Minnesota, Twin Cities
  Department of Design Innovation}
  \city{Minneapolis}
  \state{Minnesota}
  \country{USA}
}
\email{barba087@umn.edu}

\author{Ji Youn Shin}
\affiliation{%
  \institution{University of Minnesota, Twin Cities
  Department of Design Innovation}
  \city{Minneapolis}
  \state{Minnesota}
  \country{USA}
}
\email{shinjy@umn.edu}

\author{Gurumurthy Hiremath}
\affiliation{%
  \institution{University of Minnesota, Twin Cities
  Department of Pediatrics}
  \city{Minneapolis}
  \state{Minnesota}
  \country{USA}
}
\email{hiremath@umn.edu}

\author{Carlye Anne Lauff}
\affiliation{%
  \institution{University of Minnesota, Twin Cities
  Department of Design Innovation}
  \city{Minneapolis}
  \state{Minnesota}
  \country{USA}
}
\email{carlye@umn.edu}

\renewcommand{\shortauthors}{Irene Zeng, Neda Barbazi, Ji Youn Shin, Gurumurthy Hiremath, \& Carlye Anne Lauff}

\begin{abstract}
Children with congenital heart disease (CHD) often face challenges that require them to understand complex medical information from an early age in order to support lifelong care and improve health outcomes. However, prior research has rarely included young children in designing and evaluating digital tools to support health education using developmentally appropriate strategies. This study is part of a multi-phase research involving participatory design (PD), user testing, and iterative development. We present the design and refinement of a digital application that introduces basic information about CHD, including heart anatomy and healthy habits, through metaphor-based gameplay. User testing sessions with 30 children informed the redesign of interactive activities aligned with specific health conditions. Findings highlight usability, engagement, and comprehension outcomes and reveal design opportunities for supporting health literacy through serious game (SG) principles. These results inform the next phase, including further testing, refinement, and deployment in home and clinical settings.
\end{abstract}

\ccsdesc[500]{Human-centered computing}
\ccsdesc[500]{Human-centered computing~Human computer interaction (HCI)}
\ccsdesc[500]{Human-centered computing~HCI design and evaluation methods}

\keywords{Congenital Heart Disease, Serious Game, Pediatric Patients, Health Education, Patient-Centered Care, Design, User Testing}


\maketitle

\section{Introduction}
Children with chronic conditions, such as asthma, diabetes, and congenital heart disease (CHD), face ongoing challenges requiring long-term care and support \cite{barbazi_exploring_2025, cash_hank_2025, cha_its_2023, su_hidden_2024, burns_health_2022}. Among these conditions, CHD is the most common birth defect globally, affecting nearly 40,000 newborns annually in the United States \cite{cdc_data_2024}. It profoundly impacts children and families, requiring continuous medical attention and health education \cite{barbazi_developing_2025, cdc_data_2024}. This often includes multiple surgeries, frequent hospital and clinic visits, and ongoing care routines that can be difficult for young children to understand or communicate about with caregivers and providers. Despite these challenges, most educational resources are primarily aimed at parents, leaving a significant gap in child-focused health education \cite{burns_health_2022, rodts_health_2020}. Many studies focus on the parents’ experiences \cite{nikkhah_designing_2022}, such as receiving inadequate information from cardiologists \cite{arya_parents_2013} or managing caregiving-related anxiety \cite{zhang_effect_2021}. However, less attention has been given to the needs of children with CHD, including a lack of research and age-appropriate materials explicitly designed for them \cite{barbazi_developing_2025, barbazi_exploring_2025, rodts_health_2020}. Without accessible information, young patients struggle to understand their diagnosis, leading to confusion, anxiety about medical procedures, and feelings of isolation. Studies have shown that age-appropriate health education can reduce emotional distress and foster greater participation in care \cite{chong_childrens_2018}. 

Technology offers promising tools to address this gap \cite{jeong_using_2009, krasovsky_hybrid_2024, nunes_self-care_2015, seo_learning_2021, shin_identifying_2020, shin_towards_2019, thabrew_co-design_2018, whitehead_report_2024}. When integrated into educational tools, digital applications provide substantial benefits, with user experience playing a critical role in how children engage with and comprehend their health \cite{tileng_usability_2024}. Technology-mediated tools have already improved health outcomes and engagement for children with various conditions \cite{sarasmita_digital_2024, verhalen_once_2024}. For instance, “T1D Buddy”, a physical teddy bear paired with an educational app, enabled children with Type 1 diabetes understand their condition, stay engaged, and build self-management confidence \cite{khan_t1d_2023}. Similarly, an emoji-based app enabled autistic children to express emotions and communicate more effectively across neurodiverse and neurotypical groups \cite{wang_position_2024}. These examples highlight the potential of interactive, age-appropriate technologies to support children’s learning, emotional needs, and health engagement \cite{tileng_usability_2024, verhalen_once_2024}.

As part of a multi-phase research \cite{barbazi_developing_2025, barbazi_exploring_2025}, this study presents the development of a co-designed digital application (“Octo’s Heartland”) that supports health literacy and emotional engagement for children with CHD ages 4-10, through age-appropriate, interactive experiences. “Octo”, a hybrid educational tool at the center of this study, integrates a dinosaur character with a heart in its chest and medical-themed accessories alongside a digital application. The results of the study will inform the next iteration of the application, which will be deployed in real-world healthcare settings with children and their caregivers. In the previous phase of the study, the team developed an initial version of the app. However, this version lacked input from children and was primarily informed by secondary research and the designers’ interpretations of pediatric patients’ educational needs. To address this gap, in the current study, we focus on the iterative refinement of the app through participatory design (PD) \cite{sevon_participatory_2025} and usability testing with children \cite{andersen_considerations_2017}, incorporating insights to expand its heart-related educational content and enhance engagement. The design prioritizes clarity, interactivity, and accessibility to ensure that children can fully engage with the app’s features \cite{tileng_usability_2024}. Therefore, this work is structured around two research questions:
\begin{itemize}
    \item RQ1: How can we design a digital application that
    effectively supports children with CHD in learning about their
    hearts, including cardiac anatomy, common conditions, and holistic
    health practices?
    \item RQ2: How can insights gathered from child-engaged usability testing sessions be translated into design improvements for a digital application that applies serious game principles to enhance health literacy?
\end{itemize}

\section{Background and Related Work}
\subsection{Serious Games (SGs) for Children}
Prior research in HCI shows that interactive technologies are effective for supporting children’s health literacy, engagement, and emotional well-being, especially for those managing chronic conditions \cite{alexandridis_first_2023, liszio_pengunaut_2020, verhalen_once_2024, vermeulen_its_2022}. These tools can translate complex health information into developmentally appropriate, interactive formats that are easier for children to understand and apply. Building on this potential, serious games (SGs) are commonly used to embed educational objectives within engaging, goal-oriented gameplay experiences \cite{charlier_serious_2016, laamarti_overview_2014, peters_design_2023, ushaw_adopting_2015}. SGs are defined as games that “have an explicit and carefully thought-out educational purpose and are not intended to be played primarily for amusement” \cite{laamarti_overview_2014}. Since their introduction in the 1970s, SGs have integrated goal-oriented learning into gameplay, combining enjoyment with skill development, behavior change, and cognitive engagement \cite{behnamnia_review_2023}. They have been used across domains, including education, healthcare, and rehabilitation \cite{alotaibi_game-based_2024, behnamnia_review_2023, hoiseth_designing_2013, nicolaou_game_2023, sarasmita_digital_2024}. In early childhood education, SGs support problem-solving, memory, creativity, and attention through role-playing, puzzles, and narrative/memory-based games \cite{alotaibi_game-based_2024, xiong_serious_2024}. They also foster social and emotional growth. For example, \citet{saud_educational_2017} found that daily exposure to age-appropriate games improved empathy, reduced aggression, and reinforced social norms such as politeness and cooperation among kindergarten students.

In pediatric care, for instance, \citet{liszio_pengunaut_2020} demonstrated that a VR-based serious game reduced pre-MRI anxiety by allowing children to simulate the procedure in a playful, low-stakes environment. \citet{alexandridis_first_2023} developed a collaborative digital game that addressed loneliness among children with chronic illnesses by fostering shared emotional experiences. 
\citet{vermeulen_its_2022} co-designed a board game that supported disease-related conversations between children and their peers, enhancing mutual understanding and reducing social isolation. Beyond emotional support, SGs also facilitate therapeutic engagement \cite{durango_using_2015, henschke_developing_2012}. One study found that children with speech disabilities were more motivated to practice word exercises when presented through interactive, game-based formats \cite{nicolaou_game_2023}. SGs have also shown effectiveness in improving health literacy for children with conditions such as type 1 diabetes, autism spectrum disorder, and learning disabilities. These tools offer structured, developmentally appropriate experiences that promote participation in care and long-term self-management \cite{khan_t1d_2023, kim_game-driven_2023, nicolaou_game_2023}. By making abstract medical concepts more tangible and engaging, SGs promote both comprehension and confidence. They provide safe, interactive spaces where children can explore, ask questions, and rehearse health-related scenarios. For example, “Re-Mission”, a game for adolescents with cancer, enables players to control a nanobot that destroys cancer cells using chemotherapy ammunition, making treatment concepts more accessible and empowering \cite{kato_video_2008, uluhan_effects_2024}. The “GOO Platform” monitors children’s emotional responses during gameplay to support adherence in pediatric oncology \cite{damasceno_serious_2024}, while “Emotion Adventure” helps children with autism develop empathy and communication skills through interactive, narrative-based challenges \cite{kim_game-driven_2023}. These examples show the potential of serious games in pediatric care, but their impact depends on thoughtful, child-centered design.

\subsection{Game-Based Learning Design Principles for Children}

Designing educational applications for children requires careful attention to their developmental needs and learning styles, which differ significantly from adults \cite{asadzadeh_serious_2024, hoiseth_designing_2013}. This study applied four key principles from the research of \citet{kucher_principles_2021} on digital game-based learning to align with children’s cognitive abilities and emotional needs in Octo’s app design \cite{kucher_principles_2021}. 1) Meaningful \textbf{interactivity} boosts children’s learning through hands-on exploration and experimentation. 2) \textbf{Immersiveness} achieved via multisensory elements like animations and role-play, enhances motivation and engagement. 3) Effective \textbf{feedback mechanisms}, such as levels, points, and messages help children track progress. Positive feedback motivates, while explanatory feedback facilitates problem solving \cite{pan_effects_2022}. 4) \textbf{Freedom of exploration} allows children to experiment without fear of failure, promoting confidence and persistence through mistakes. In addition to these strategies, aligning content with children’s cognitive strengths—such as visual, interactive, and non-verbal elements like color and movement—supports attention and retention without cognitive overload \cite{asadzadeh_serious_2024, kucher_principles_2021, valenza_serious_2019}.

Serious games have improved health literacy, self-management, and emotional well-being in children and adolescents with chronic conditions such as diabetes, cancer, and asthma \cite{charlier_serious_2016, delmas_serious_2017, kato_video_2008, sarasmita_digital_2024}. However, younger children, particularly those under age 7, and children with special needs are often excluded from designing and evaluating these tools \cite{buckmayer_participatory_2024, guha_designing_2008}. This exclusion results from developmental, methodological, and ethical challenges. Developmentally, young children may struggle with abstract reasoning, sustained attention, and verbal expression, limiting standard usability testing \cite{andersen_considerations_2017, banker_usability_2022, hoiseth_designing_2013, vang_reflections_2024}. Methodologically, most participatory design frameworks target older children and are rarely adapted for early childhood or children with complex needs \cite{buckmayer_participatory_2024, constantin_pushing_2019, druin_cooperative_1999, druin_role_2002, frauenberger_designing_2011, lindberg_participatory_2013, stalberg_childs_2016}. Studies emphasize the need for specialized scaffolding strategies, such as breaking tasks into smaller steps, providing visual or verbal prompts, and gradually reducing support, to enable meaningful participation in co-design \cite{guedes_scaffolding_2024, druin_cooperative_1999, druin_role_2002, stalberg_childs_2016}. Ethically, concerns about consent, vulnerability, and institutional hesitancy further restrict research involving this age group \cite{mcnally_childrens_2016, sevon_participatory_2025}. As a result, few SGs reflect the developmental needs or lived experiences of young children and, to our knowledge, none focus specifically on children with CHD \cite{barbazi_exploring_2025}. Given the complex care demands of CHD and its developmental impact on young patients \cite{burns_health_2022, rodts_health_2020}, Octo was designed to address this gap through interactive modules that integrate cardiac education with emotional support. These modules support children across a range of developmental stages, covering core topics such as basic heart anatomy, common CHD conditions, clinic visit preparation, health habits, and emotional regulation. The Octo project addresses this gap by applying SG principles to CHD, aiming to promote health literacy, emotional resilience, and child-led participation through co-designed, play-based education.

\section{Preliminary User Testing}

\begin{figure}[h]
  \centering
  \includegraphics[width=\linewidth]{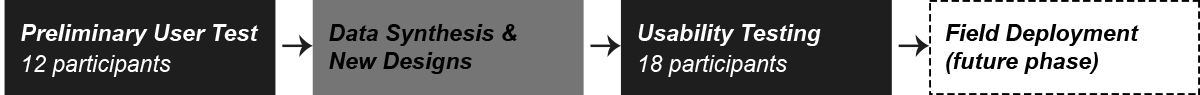}
  \caption{Research and Design Process of Octo Digital App.}
  \Description{Research and Design Process of Octo Digital App.}
  \label{fig:Timeline}
\end{figure}

This research applies a PD approach, which emphasizes collaboration with users throughout the design process to ensure outcomes reflect their needs and expectations \cite{druin_cooperative_1999, druin_role_2002, sevon_participatory_2025}. PD is well-suited for this project, as children with CHD benefit from developmentally appropriate, child-led tools that enhance engagement and retention \cite{kucher_principles_2021}. To integrate children’s perspectives into SG design \cite{valenza_serious_2019}, the PD process included observational research and usability testing at Camp Odayin, a medically supported, community-based program for families affected by heart disease \cite{CampOdayin_About}. Rather than treating children as passive testers, the study positioned them as active informants—participants who shaped design through independent exploration, in-situ reactions, and responses to open-ended questions \cite{buckmayer_participatory_2024, druin_cooperative_1999, druin_role_2002, sevon_participatory_2025,yip_examining_2017,qi_participatory_2025}. To accommodate developmental differences, we employed observational methods over task-based approaches for younger children, attending to non-verbal cues such as hesitation, laughter, or repeated actions. For older participants, we used verbal prompts and structured follow-up questions to elicit feedback specific to the design. Across both age groups, we documented usability patterns, frustrations, and emotional responses, which were thematically coded to inform refinements in navigation, pacing, activity structure, and feedback features. The preliminary study and usability testing received IRB approval from the University of Minnesota (STUDY00020670). Figure~\ref{fig:Timeline} illustrates the iterative research and design process, including multiple rounds of observation, synthesis, testing, and refinement.

\subsection{Testing Context and Procedure}

\begin{figure}[h]
  \centering
  \includegraphics[width=\linewidth]{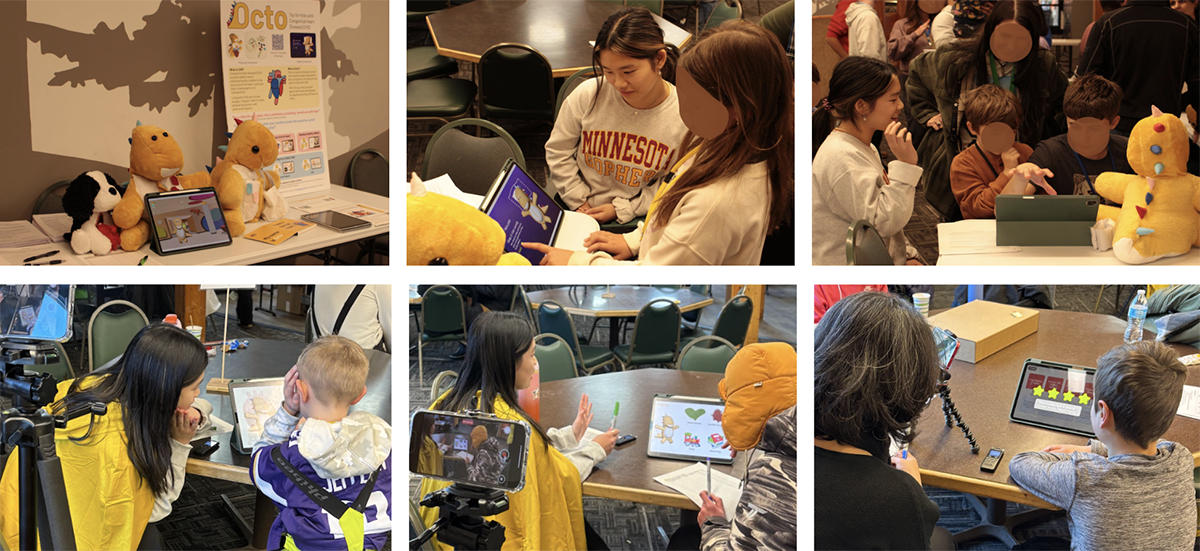}
  \caption{Camp Odayin Preliminary Sessions (October 19th, 2024) and Usability Testing Sessions (February 16th, 2025).}
  \Description{Camp Odayin Preliminary Sessions (October 19th, 2024) and Usability Testing Sessions (February 16th, 2025).}
  \label{fig:UserTest}
\end{figure}

A preliminary study was conducted in-person on October 19, 2024 at Camp Odayin. The research team gathered insights from children with CHD by observing their interactions with the existing Octo digital application prototype. The objective was to identify effective elements, uncover usability challenges, and evaluate the influence of the educational content, while noting participants’ reactions to the app. Twelve children participated: 9 in the target age range of 4–10 years (2 aged 4, 2 aged 6, 2 aged 7, 2 aged 8, 1 aged 10), and 3 older participants (ages 12, 13, and 16). Six were male and six female. While the focus was on ages 4–10, older children were intentionally included for their lived experience with CHD, offering insights younger children might not express. The testing protocol followed principles from \citet{andersen_considerations_2017}, accommodating short attention spans (<30 min) and relying on non-verbal cues, as children under 12 may struggle with think-aloud practices. Given participants’ young ages, testing emphasized observation over task-based evaluation \cite{andersen_considerations_2017, read_childcomputer_2013}. Each session lasted around 15 min., with handwritten notes taken during and after. Children were introduced to both the Octo plush toy and app to create a familiar environment before the testing process. The researchers encouraged independent exploration of the app and observed navigation, noting frustration, enjoyment, and engagement patterns. At the end of each session, participants verbally answered five open-ended questions about the app’s design and their experience with the prototype (e.g., \textit{What would make this ‘adventure’ [activities] better?}) (Figure~\ref{fig:UserTest}).

\subsection{Preliminary Testing Outcomes}

Handwritten notes by the lead researcher, including observations and quotes, were thematically coded \cite{braun_thematic_2012} using Miro, a digital collaboration platform. Coding focused on usability (e.g., activity navigation), educational content (e.g., game instructions), and play reactions (e.g., emotional engagement, confusion). Codes were organized in two stages: first, to identify high-level themes; and second, to create subthemes. There were seven high-level themes and eighteen total subthemes: 1) navigation (audio preferences, buttons, and navigation between activities); 2) instructions (directions not guiding, feedback); 3) emotions/reactions (excitement, feeling represented, helpful for younger patients, helpful for siblings); 4) pace (pace was inefficient, pace was efficient); 5) technical issues (glitching components, sensitivity issues, misleading design); 6) questions (cater to diagnosis, when to use); 7) aesthetics/design (reaction, request). For example, codes such as ‘\textit{Amount of information per screen is digestible}’ (observation) and “\textit{The pace felt just right}” (7 yrs) were grouped under the high-level theme of ‘pace’ and further categorized under the subthemes ‘effective pacing’ and ‘ineffective pacing’.

\textbf{Usability findings:} Several participants struggled with inconsistent navigation and unintuitive interactions. Some buttons lacked call-to-actions, and the ‘Next’ buttons were excessive and inconsistently placed. One child (7 yrs) noted, “\textit{I’m not sure what I’m supposed to do right now},” pointing to unclear instructions. Despite these issues, younger users showed strong engagement with the interface, one exclaimed, “\textit{I want to keep playing!}” (4 yrs), suggesting appeal despite usability issues. In terms of activity pacing, older participants often felt restricted by the rigid game structure and step-by-step play, while younger ones stated that “\textit{the pace feels just right}” (7 yrs).

\textbf{Educational and Emotional findings:} The educational content (e.g., clinic simulation of EKG and blood pressure activities) effectively engaged participants but could benefit from more elaborate interactions. The content was positively received, with one older participant reflecting, \textit{“This [app] would’ve been so helpful to have when I was younger!”} Additionally, participants appreciated their experiences being represented in the content, one child said, \textit{“[Octo] looks just like me!”} (5 yrs), highlighting inclusive design in strengthening learning connections. However, both younger and older participants expressed the desire for more interactive and narrative-driven content.

These outcomes informed design criteria for future work, emphasizing the need for intuitive navigation, consistent and accessible elements, and a balance between structured learning and playful exploration. Future designs should prioritize clear guidance through labeled components and call-to-actions (CTAs), adaptable activity pacing, and an engaging yet approachable experience. Furthermore, character design is integral to forming an emotional connection and excitement.

\begin{table*}
  \caption{Comparison table of the four gamification approaches designed for usability testing during February 2025.}
  \label{tab:Method}
  \begin{tabular}{L{1.04in}L{1.4in}L{1.04in}L{1.04in}L{0.6in}L{1.04in}} 
    \toprule
    Activity Name &CHD Condition & Interactivity & Immersiveness & Exploration & Feedback\\
    \midrule
    1. Heart Island	& Ventricular Septal Defect & Problem-solving & Nature-inspired metaphors & Flexible structure	& Messages (positive + constructive) \\
    \\
    2. Train	& Pulmonary Stenosis	& Problem-solving \& puzzle	& Train adventure with 8 stops	& Linear structure	& Messages (positive + constructive) \\
    \\
    3. Cardia Kingdom	& Coarctation of the Aorta	& Problem-solving \& scavenger hunt	& Medeival/magic narrative	& Flexible structure	& Messages (positive + constructive) \\
    \\
    4. Matching Game	& VSD, Pulmonary Stenosis, Coarctation of the Aorta, ASD	& Cardiac terminology activities	&“Quiz”-like knowledge game	&Both	&Color (right/wrong); stars (reward)\\
    \bottomrule
  \end{tabular}
\end{table*}

\section{Usability Testing}

\subsection{Usability Testing Setting and Methodology}

Usability testing assesses whether a product is effective, efficient, and easy to learn \cite{andersen_considerations_2017, tileng_usability_2024}. The primary objective was to evaluate four new, distinct activities to determine which elements effectively conveyed educational content and implemented game-based learning design principles: interactivity, immersiveness, exploration, and feedback \cite{kucher_principles_2021}. Each activity featured a different narrative theme, game structure (e.g., linear sequence vs. exploratory), CHD condition introduced, and sequence of content presented (e.g., game followed by education or vice versa) as described in Table~\ref{tab:Method}. Eighteen children with CHD participated (ages 6–16): 1 aged 6, 1 aged 7, 3 aged 8, 2 aged 9, 3 aged 10, 1 aged 11, 1 aged 13, 3 aged 14, 2 aged 15, and 1 aged 16. Ten were male and eight female. The protocol mirrored the preliminary study, emphasizing observation and age-appropriate methods \cite{andersen_considerations_2017}. Participants worked alone or in pairs to encourage idea sharing. Sessions were video recorded to capture verbal and non-verbal feedback (e.g., facial expressions). Each 15-minute session began with two researchers collecting demographics and introducing testing goals. Participants explored the four activities in any order. We observed which activities they chose, usability, moments of excitement and frustration, and CHD content comprehension. After each activity, participants answered four short questions on likes/dislikes, learning, and suggestions (e.g., \textit{How can we make the game better for kids with heart problems?}). After all activities, they answered six questions on emotional experience, favorite activity, difficulty, and heart knowledge (e.g., \textit{Which game was your favorite and Why?}) (Figure~\ref{fig:UserTest}).

\subsection{Usability Testing Outcomes}

Audio recordings from usability testing sessions were transcribed verbatim for thematic data analysis \cite{braun_thematic_2012}. Transcripts were coded sentence by sentence to identify patterns in engagement, feedback, and educational comprehension. Video observations were also coded. Codes were transferred to Miro and color-coded by session. Coding occurred in two stages: first, to identify high-level usability and design themes; second, to extract activity-specific insights. This analysis informed refinements to better meet children’s needs. Participants responded positively to gamifying CHD content; one child noted, “\textit{I’m happy you were able to make this [CHD] into a game.}” Yet knowledge gaps emerged: one participant said, “\textit{I don’t know much about anything}” during the matching game, underscoring the need for stronger CHD education. \textbf{Heart Island Outcomes (Activity 1):} Nature metaphors were well received, but some children struggled to connect them to medical content. The problem-solving was too easy, and users desired more challenge. The flexible structure reduced restrictions, though navigation needed improvement. \textbf{Train Outcomes (Activity 2):} The theme was visually engaging but needed clearer links to medical concepts. The puzzle format stood out; many chose this as their favorite and requested more levels. \textbf{Cardia Kingdom Outcomes (Activity 3):} The activity was intuitive and the narrative engaging, but participants wanted more immediate feedback when making choices in the “tool-box”. The river metaphor was understandable and made the anatomical information more digestible. \textbf{Matching Outcomes (Activity 4):} Engagement varied based on individual preferences. Some participants found viewing words instead of images overwhelming, particularly those with limited prior knowledge of the content. While the color-coded feedback was helpful, explanatory feedback may enhance comprehension. Furthermore, the game's repetitive structure would benefit from incorporating progressive difficulty.

Connecting back to the game-based learning design principles \cite{kucher_principles_2021}, the usability testing findings resulted in design implications. For interactivity, problem-solving based activities like the train puzzle were the most appealing. For immersiveness, nature-inspired metaphors (island, river, etc.) were best-received. For exploration, a blend of flexible/linear game structure was most engaging for children. For feedback, encouraging positive messages and explanatory messages effectively guided players throughout the activities.

In terms of usability, participants commonly mentioned the need for more activity guidance, narrative cohesion, and rewards to support engagement and understanding. Overall, the activities showed strong potential to educate and engage, but need refinement to improve clarity, feedback, and conceptual framing.

\section{Result}

\subsection{Design Criteria Based on Types of CHD}
Based on the usability testing insights, design criteria were refined to guide the final design process. The criteria were divided into two categories: 1) usability and 2) educational and emotional outcomes. Usability criteria involve intuitive navigation (e.g., consistent call-to-action buttons), flexible pacing (e.g., ability to exit activities or explore others), and accessible design (e.g., readability, clear visual cues). Educational outcomes involve storytelling or problem-solving to introduce heart anatomy, medical terminology, CHD conditions, and relevant health practices. Emotional outcomes refer to maintaining a supportive and uplifting tone to prevent users from feeling negative emotions like shame. The final design includes four CHD conditions: Coarctation of the Aorta, Pulmonary Stenosis, Ventricular Septal Defect (VSD), and Atrial Septal Defect (ASD). Coarctation of the Aorta and Pulmonary Stenosis represent two narrowing conditions, while VSD and ASD represent two opening conditions. They were selected for their frequency in CHD and ability to showcase key concepts of anatomy, function, and impact.

\subsection{Final Design Features}
\begin{figure}[h]
  \centering
  \includegraphics[width=\linewidth]{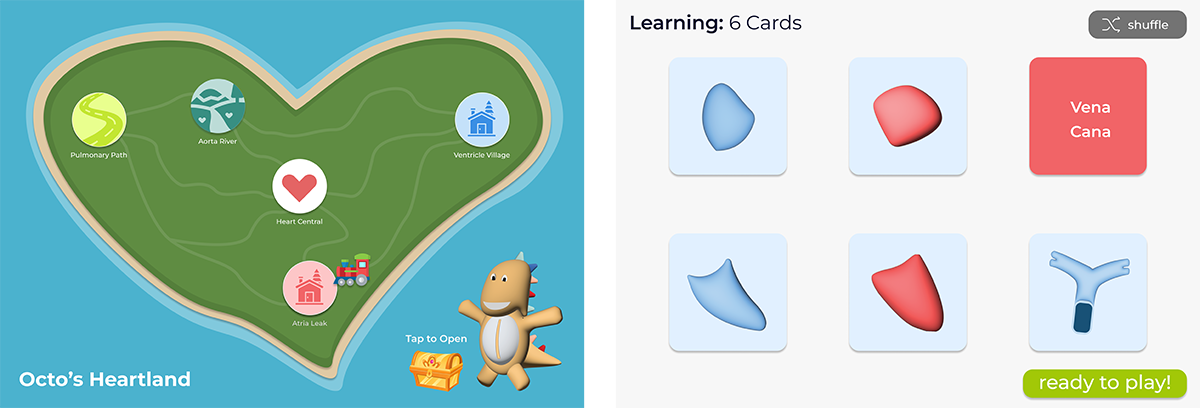}
  \caption{Octo's Heartland Home Screen (left); Heart Central Learning Cards (right).}
  \Description{Octo's Heartland Home Screen; Heart Central Learning Cards.}
  \label{fig:Heart Central Learning Card}
\end{figure}

The final design centers around “Octo’s Heartland,” a heart-shaped island navigated by train with five interactive stops. Children begin at \textit{Heart Central}, where they explore heart anatomy, learn fun facts, and ease into educational content. This section gradually increases in difficulty, supporting freedom of exploration through non-linear progression, which helps maintain motivation \cite{kucher_principles_2021}. Each stop represents a CHD condition through three levels that blend metaphor, interactivity, and learning (Figure~\ref{fig:Heart Central Learning Card}). 

\begin{figure}[h]
  \centering
  \includegraphics[width=\linewidth]{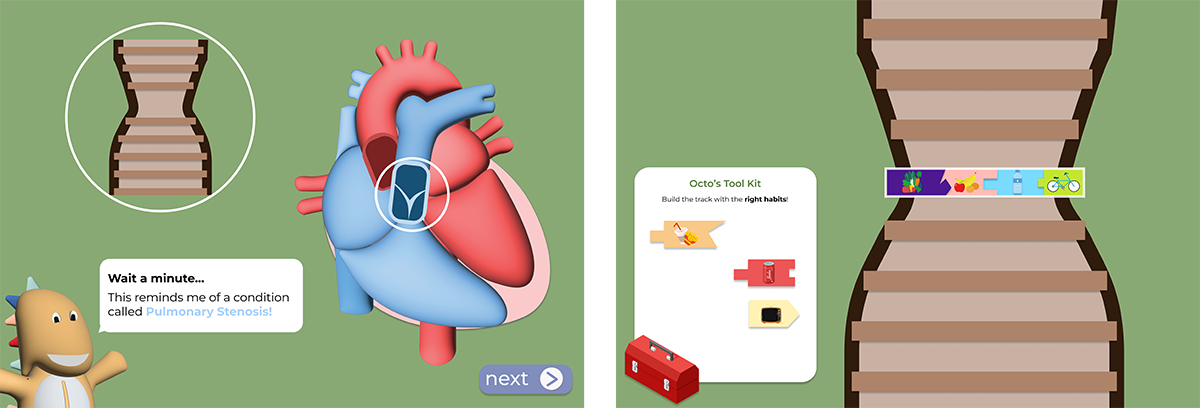}
  \caption{Explaining Track Metaphor (left); Pulmonary Pathway Level 2 (building habits) (right).}
  \Description{Explaining Track Metaphor; Pulmonary Pathway Level 2 (building habits).}
  \label{fig:Track Metaphor and Pulmonary Pathway}
\end{figure}

In the \textit{Pulmonary Pathway}, children in level 1 engage with a train track metaphor for pulmonary stenosis, where they widen tracks using blocks. This interactive approach allows for hands-on exploration. In level 2, players select heart-healthy habits to ease the narrowing, reinforcing positive behavioral feedback. By level 3, they simulate treatment using a balloon metaphor, enhancing immersiveness and reinforcing learning through role-play  (Figure~\ref{fig:Track Metaphor and Pulmonary Pathway}).

\begin{figure}[h]
  \centering
  \includegraphics[width=\linewidth]{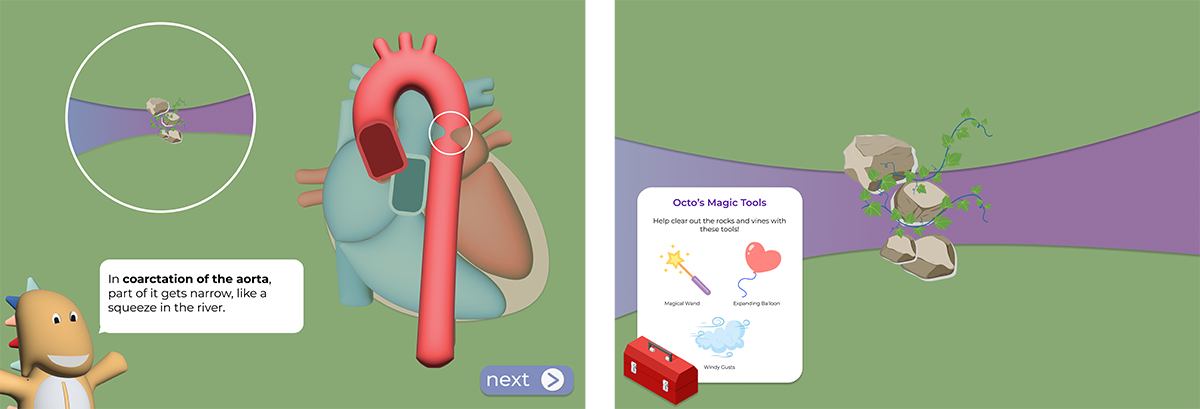}
  \caption{Explaining River Metaphor (left); Aorta River Level 3 (tool-kit to clear river) (right).}
  \Description{Explaining River Metaphor; Aorta River Level 3 (tool-kit to clear river).}
  \label{fig:River Metaphor and Aorta River}
\end{figure}

The \textit{Aorta River} mirrors coarctation of the aorta by having children widen a narrowed section of the river, promoting interactive experimentation and problem-solving. In level 1, players are introduced to the river narrowing and collect tools to clear the river blockage (e.g., vines, rocks). In level 2, children prepare Octo with healthy choices that fill a heart progress bar, linking habits to health. The final level simulates treatment by widening the river, reinforcing active learning through engagement and feedback \cite{kucher_principles_2021} (Figure~\ref{fig:River Metaphor and Aorta River}).

\textit{Ventricle Village} introduces VSD using a house metaphor. Players locate an opening between rooms in level 1, and they help Octo feel better by making decisions related to self-care in level 2. In level 3, players patch the wall, reinforcing problem-solving and understanding the medical condition. Similarly, \textit{Atria Leak} uses the upstairs rooms of a house to represent atrial defects, with children sorting objects to seal openings in the wall. These interactive elements build up emotional engagement and foster understanding  (Figure~\ref{fig:House Metaphor and
Restoring House}). 

\begin{figure}[h]
  \centering
  \includegraphics[width=\linewidth]{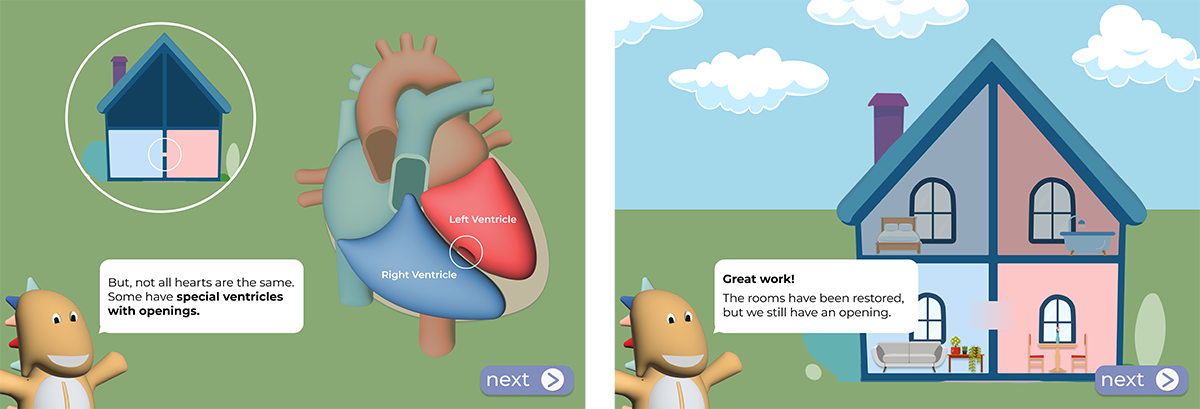}
  \caption{Explaining House Metaphor ("downstairs") (left); Restoring House (clear mixing items) (right).}
  \Description{Explaining House Metaphor ("downstairs"); Restoring House (clear mixing items).}
  \label{fig:House Metaphor and
Restoring House}
\end{figure}

The activities weave anatomical learning into gameplay, creating an immersive experience with visual and interactive elements. Users earn stars for each level, reinforcing progress, creating motivation, and balancing education and play.

\section{Discussion, Conclusion, and Future Work}

The redesigned digital application builds on previous versions to support children with congenital heart disease by expanding educational offerings, incorporating holistic health practices, and deepening understanding of CHD conditions \cite{barbazi_developing_2025, barbazi_exploring_2025, walsh_layered_2010}. Through a novel adventure-like narrative centered around Octo, the updated design allows users to navigate activities in a more flexible and engaging manner. By implementing three levels per activity, the app encourages gradual learning, sustained motivation, and a sense of accomplishment, reflecting multilevel structures found effective in serious games designed for primary school children \cite{ghisio_multimodal_2017}. 

This study contributes to HCI and CSCW communities by highlighting how PD and SG principles can be adapted to support learning among young children with complex needs \cite{buckmayer_participatory_2024, hoiseth_designing_2013, verhalen_once_2024}. Conducted in a medically supported camp setting, our sessions engaged children as young as 4 in a structured yet playful environment that allowed for authentic feedback. Rather than relying on traditional task-based protocols, we used open-ended app exploration, paired with verbal prompts and observational data collection, to gather insights across diverse developmental levels \cite{andersen_considerations_2017, banker_usability_2022, barbazi_developing_2025, kucher_principles_2021, vang_reflections_2024}. Children interacted with a high-fidelity prototype featuring multiple narrative scenarios, and their reactions, both verbal and non-verbal, were used to evaluate usability, emotional engagement, and conceptual understanding \cite{banker_usability_2022, braun_thematic_2012, laamarti_overview_2014}. These methods enabled the inclusion of children often excluded from HCI due to age or medical vulnerability \cite{asadzadeh_serious_2024, durango_using_2015, kucher_principles_2021}. In parallel, metaphorical game design (e.g., clearing a river to represent treatment) translated complex health content into accessible, emotionally resonant experiences \cite{charlier_serious_2016, delmas_serious_2017, verhalen_once_2024}. This work offers applicable strategies for designing interactive systems in any domain where abstract content must be communicated to young or neurodiverse users, not only in pediatric care, but also in education, therapy, and public communication. This research advances inclusive HCI practices by showing how to meaningfully involve children in the design and evaluation of digital tools \cite{buckmayer_participatory_2024, druin_cooperative_1999, druin_role_2002, frauenberger_designing_2011, guha_designing_2008, sevon_participatory_2025}.

Although Octo is designed primarily for children, it also alleviates the educational burden on parents by enabling children to independently understand their health condition. This reduces the need for parents to constantly interpret medical information. Additionally, by helping children grasp these concepts, Octo enhances communication between families and healthcare providers, fostering better coordination and more efficient care \cite{barbazi_exploring_2025, burns_health_2022, rodts_health_2020, hong_triggerhunter_2010}.

This phase represents the digital component of Octo and plays a central role in advancing the hybrid intervention. Building on our prior research with children, parents, and providers \cite{barbazi_developing_2025, barbazi_exploring_2025}, it addresses a key gap in the prototype: the need for more engaging and developmentally appropriate heart-related content. This phase strengthens Octo’s ability to support pediatric health literacy and aligns with stakeholder priorities by testing and refining multiple narrative-based activities. The findings guide improvements in usability and content design, positioning the app as an interactive component of Octo’s hybrid model. Future work will broaden the scope of CHD conditions represented, implement verbal instructions, and conduct further usability testing to assess the new activity structure and its effectiveness. These efforts will support the transition from a conceptual prototype to a fully functional application that can be deployed in home and clinical settings.

\begin{acks}
The authors would like to thank all the participants and staff at Camp Odayin who made this study possible. Appreciation to past students: Jessica Jenkins Espinosa, Jonathan Jakubas, Grace Rubas, Levi Skelton, and Andy Thai. This research was supported by the Lasting Imprint Foundation, the Office of Discovery and Translation (ODAT)/Pediatric Device Innovation Consortium (PDIC) grant, and the University of Minnesota’s Research Opportunities Program (UROP).
\end{acks}

\bibliographystyle{ACM-Reference-Format}
\bibliography{sample-base}

\end{document}